\documentclass{webofc}
\usepackage[varg]{txfonts}   

\newcommand{\ccbar}{{$\text{c} \overline{\text{c}}$ }\xspace}
\newcommand{\accbar}{{$\langle \text{c} \overline{\text{c}}\rangle$ }\xspace}
\newcommand{\AGeVc}{\ensuremath{A\,\mbox{Ge\kern-0.1em V}\!/\kern-0.08em c}\xspace}
\newcommand{\AGeV}{\ensuremath{A\,\mbox{Ge\kern-0.1em V}}\xspace}

\newcommand{\pip}{\ensuremath{\pi^+}}
\newcommand{\km}{\ensuremath{\textup{K}^-}\xspace}

\newcommand{\Dzero}{\ensuremath{\textup{D}^0}\xspace}
\newcommand{\Dzerobar}{\ensuremath{\overline{\textup{D}^0}}\xspace}
\newcommand{\Dplus}{\ensuremath{\textup{D}^+}\xspace}
\newcommand{\Dminus}{\ensuremath{\textup{D}^-}\xspace}

\begin{document}
\title{Open Charm measurements at the NA61 experiment at CERN SPS}
%
%

\author{\firstname{Dag}       \lastname{Larsen}  \inst{1}  \fnsep\thanks{\email{dag.larsen@cern.ch}}         \and
        \firstname{Anastasia} \lastname{Merzlaya}\inst{1,2}\fnsep\thanks{\email{anastasia.merzlaya@cern.ch}}
        for the NA61/SHINE collaboration
}

\institute{Jagiellonian University, Krakow, Poland
\and
           Saint-Petersburg State University, Saint-Petersburg, Russia
          }

\abstract{%
The strong interactions programme of the NA61/SHINE experiment at CERN SPS has been extended through the use of new silicon Vertex Detector which provides precise measurements of exotic particles with short lifetime.
The detector was designed to meet the challenges of primary and secondary vertexes reconstruction at high spatial resolution.

An initial version of the Vertex Detector called SAVD (Small Acceptance Vertex Detector) was installed last end of 2016, and data was from Pb+Pb collisions was collected in 2016, for Xe+La in 2017, as well as further Pb+Pb collisions this year. First indication of a D$^0$ peak at SPS energies has been observed.

The physics motivation behind the open charm measurements will be discussed, as well as the analysis of collected data on open charm production and the future plans of open charm measurements in NA61/SHINE experiment related to the upgraded version of the vertex detector.
}
\maketitle
\section{Introduction}
\label{intro}
The SPS Heavy Ion and Neutrino Experiment (NA61/SHINE) \cite{na61} is a fixed-target experiment located at the CERN Super Proton Synchrotron (SPS).
The NA61/SHINE detector is optimised to study hadron production in hadron-proton, hadron-nucleus and nucleus-nucleus collisions.
It consists of a large acceptance hadron spectrometer with excellent capabilities in charged particle momentum measurements and identification by a set of eight Time Projection Chambers (TPC) as well as Time-of-Flight (ToF) detectors. 
The strong interaction research program of NA61/SHINE is dedicated to study the properties of the onset of deconfinement and search for the critical point of strongly interacting matter. 
These goals are being pursued by investigating p+p, p+A and A+A collisions at different beam momenta from 13A to 150\AGeVc.
In 2016 NA61/SHINE was upgraded with the Small Acceptance Vertex Detector (SAVD) based on MIMOSA-26AHR sensors developed in IPHC Strasbourg.  
Construction of this device was mostly motivated by the importance and the possibility of the first direct measurements of open charm meson production in heavy ion collisions at SPS energies.  
Precise measurements of charm hadron production by NA61/SHINE are expected to be performed in 2022--2024. 
The related preparations have started already.  

\section{Physics motivation for open charm measurements}
\label{sec2}

\begin{figure}[]
\centering
\includegraphics[width=\textwidth]{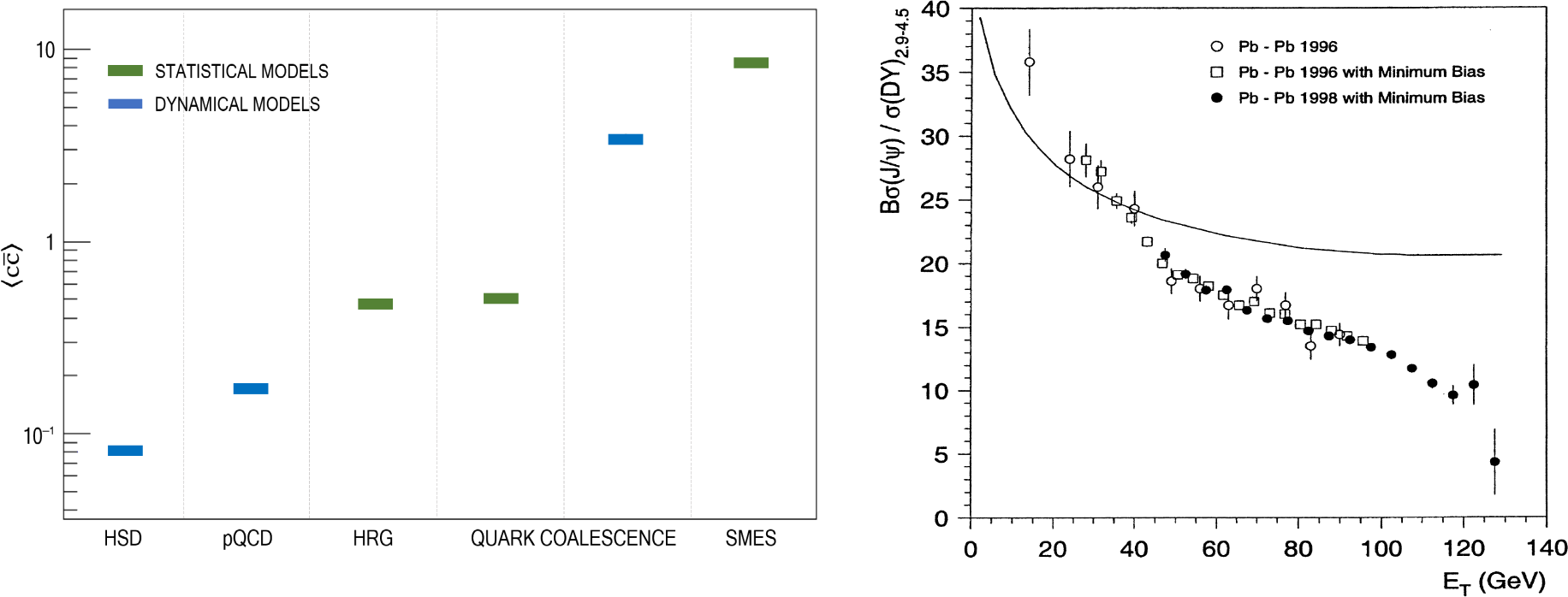}
\caption{
\textit{Left.}
Mean multiplicity of charm quark pairs produced in the full phase space in central Pb+Pb collisions at 158\AGeVc calculated with dynamical models (blue bars): HSD~\cite{Linnyk:2008hp,TSong}, pQCD--inspired~\cite{Gavai:1994gb,BraunMunzinger:2000px}, and Dynamical Quark Coalescence~\cite{Levai:2000ne}, as well as statistical models (green bars): HRG~\cite{Kostyuk:2001zd}, Statistical Quark Coalescence~\cite{Kostyuk:2001zd}, and SMES~\cite{Gazdzicki:1998vd}.
\textit{Right.}
The ratio of $\sigma_{J/\Psi}/\sigma_{DY}$ as a function of transverse energy (a measure of collision violence or centrality) in Pb+Pb collisions at 158A GeV measured by NA50. The curve represents the J/$\Psi$ suppression due to ordinary nuclear absorption~\cite{Abreu:2000ni}.}
\label{fig:models}
\end{figure}

One of the important issues related to relativistic heavy-ion collisions is the mechanism of charm production.
Several model predictions was introduced to describe charm production.
Some of them are based on the dynamical approach and some on the statistical approach.
The estimates from these approaches for average number of produced $\text{c}$  and $\overline{\text{c}}$ pairs (\accbar) in central Pb+Pb collisions at 158\AGeVc differ by up to a factor of 50 \cite{ana1,ana2} which is illustrated in Figure \ref{fig:models} (\textit{left}).

Charm mesons are of vivid interest in the context of the phase-transition between confined hadronic matter and the quark-gluon plasma (QGP).
The \ccbar pairs produced in the collisions are converted into open charm mesons and charmonia ($\text{J}/\psi$ mesons and its excited states).
The production of charm is expected to be different in confined and deconfined matter.
This is caused by different properties of charm carriers in these phases.
In confined matter the lightest charm carriers are D mesons, whereas in deconfined matter the lightest carriers are charm quarks.
Production of a DD pair (2m$_{D} = 3.7$ GeV) requires energy about 1 GeV higher than production of a \ccbar pair (2m$_{c} = 2.6$ GeV).
The effective number of degrees of freedom of charm hadrons and charm quarks is similar \cite{ana7}.
Thus, more abundant charm production is expected in deconfined than in confined matter.
Consequently, in analogy to strangeness \cite{ana2,ana8}, a change of collision energy dependence of \accbar may be a signal of onset of deconfinement.





Figure~\ref{fig:models} (\textit{right}) shows results on  $\langle \text{J}/\psi \rangle$ production normalised to mean multiplicity of Drell-Yan pairs in Pb+Pb at the top SPS energy obtained by NA50 collaboration.
The solid line shows model prediction for normal nuclear absorption of $\text{J}/\psi$ in the medium. 
NA50 observed that the $\text{J}/\psi$ production is consistent with normal nuclear matter absorption for peripheral collision and it is suppressed for more central collisions. 
This so called anomalous suppression was attributed to the  $\text{J}/\psi$ dissociation effect in the deconfined medium. 
However, the above result is based on the assumption that \accbar  $\sim \langle DY \rangle $ that may be incorrect due to many effects, such as shadowing or parton energy loss ~\cite{Satz:2014usa}.
Thus the effect of the medium on \ccbar binding can only be quantitatively determined by comparing the ratio of $\langle \text{J}/\psi \rangle$ to \accbar in nucleus-nucleus to that in proton-proton. 
However, in Pb+Pb data the onset of colour screening should already be seen in $\langle \text{J}/\psi \rangle$ to \accbar ratio centrality dependence.
This clearly shows the need for large statistic data on \accbar.

\section{Performance of SAVD}
\label{sec3}

The SAVD was built using sixteen CMOS MIMOSA-26 sensors~\cite{mimosa26}.
The basic sensor properties are:
$18.4 \times 18.4~\mu$m$^2$  pixels, 115~$\mu$s time resolution, $10 \times$ 20~mm$^2$ surface, 0.66 mega pixel, 50~$\mu$m thick.	
The estimated material budget per layer, including the mechanical support, is 0.3\% of a radiation length.
The sensors were glued to eight ALICE ITS ladders~\cite{Abelev:1625842}, which were mounted on two horizontally movable arms and spaced by 5~cm along the \textit{z} (beam) direction.
The detector box was filled with He (to reduce beam-gas interactions) and contained an integrated target holder to avoid unwanted material and multiple Coulomb scattering between target and detector.
The first test of the device was performed in December 2016 during a Pb+Pb test run. 
The test allowed to demonstrate: tracking in a large track multiplicity environment, precise primary vertex reconstruction, TPC and SAVD track matching and and allowed to make a first search for the \Dzero and \Dzerobar signals.

Tracking is rather challenging due to the inhomogeneous magnetic field~\cite{Merzlaya:2017nvb}.
As track model a parabola in the x-z plane, and a straight line in y-z plane was used~\cite{Merzlaya:2018uta}.

\begin{figure}[]
\centering
\includegraphics[width=\textwidth]{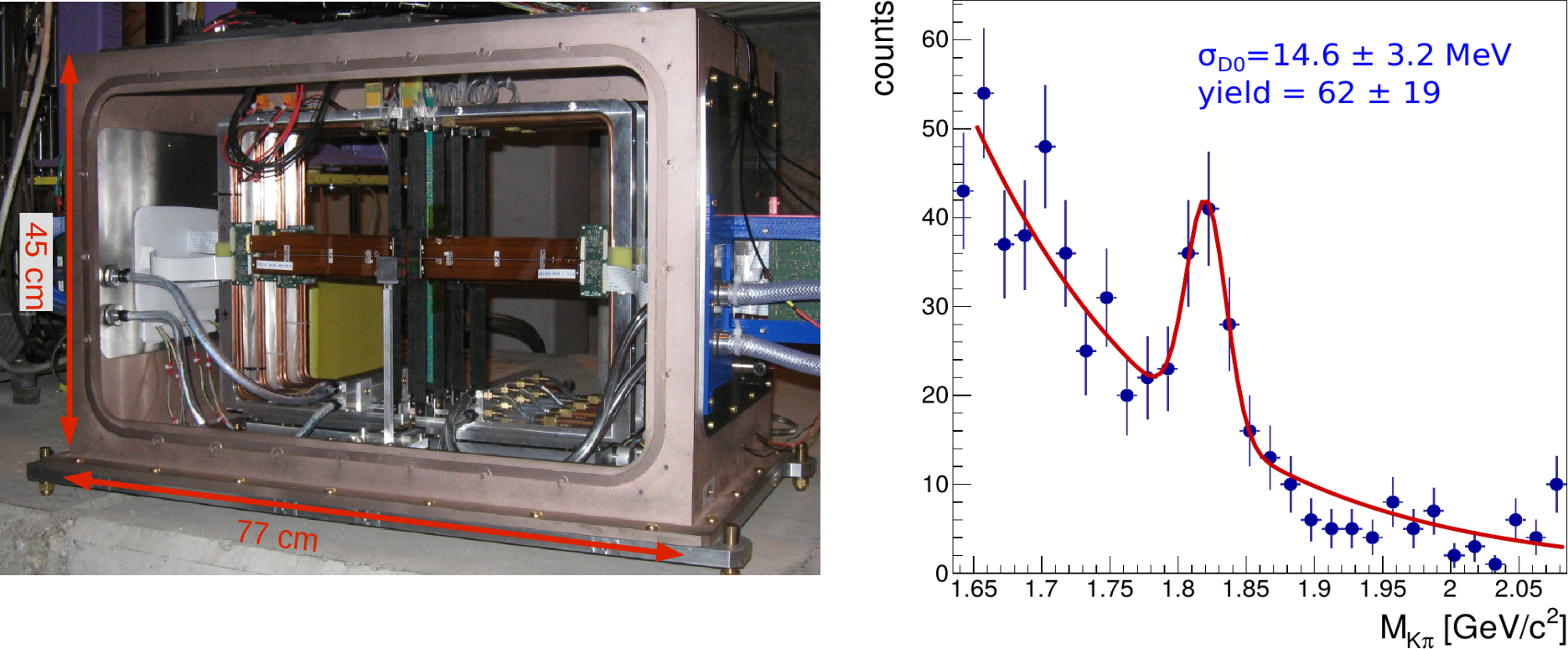}
\caption{
\textit{Left.}
The SAVD used by NA61/SHINE during the data taking in 2016 and 2017. 
\textit{Right.}
The invariant mass distribution of \Dzero and \Dzerobar candidates in central Pb+Pb collisions at 150\AGeVc after the background suppression cuts.
The particle identification capability of NA61/SHINE was not used at this stage of the analysis~\cite{ana1}.
}
\label{fig:firstD0}
\end{figure}

Based on these data, the spatial resolution of the SAVD was determined. 
Cluster position resolution is $\sigma_{\text{x,y}}(\text{Cl}) \approx 5~\mu\mathrm{m} $ and primary  vertex resolution in the transverse plane is $\sigma_{\text{x}}(\mathrm{PV}) \approx  5~\mu\mathrm{m} $, $\sigma_{\text{y}}(\text{PV}) \approx 1.8~\mu\mathrm{m} $\footnote{$\sigma_{\text{x}}(\text{PV}) > \sigma_{\text{y}}(\text{PV})$ because $B_\text{y} \gg B_\text{x}$}, and along the beam direction is $\sigma_{\text{z}}(\text{PV}) \approx 30\mu\mathrm{m} $ for a typical multiplicity of events recorded in 2016. 
The obtained primary vertex resolution along the beam direction of 30~$\mu$m was sufficient to perform the search for the \Dzero and \Dzerobar signals. 
Figure~\ref{fig:firstD0} (\textit{right}) shows the first indication of a \Dzero and \Dzerobar peak obtained using the data collected during the Pb+Pb run in 2016.

\begin{figure}[]
\centering 
\includegraphics[width=\textwidth]{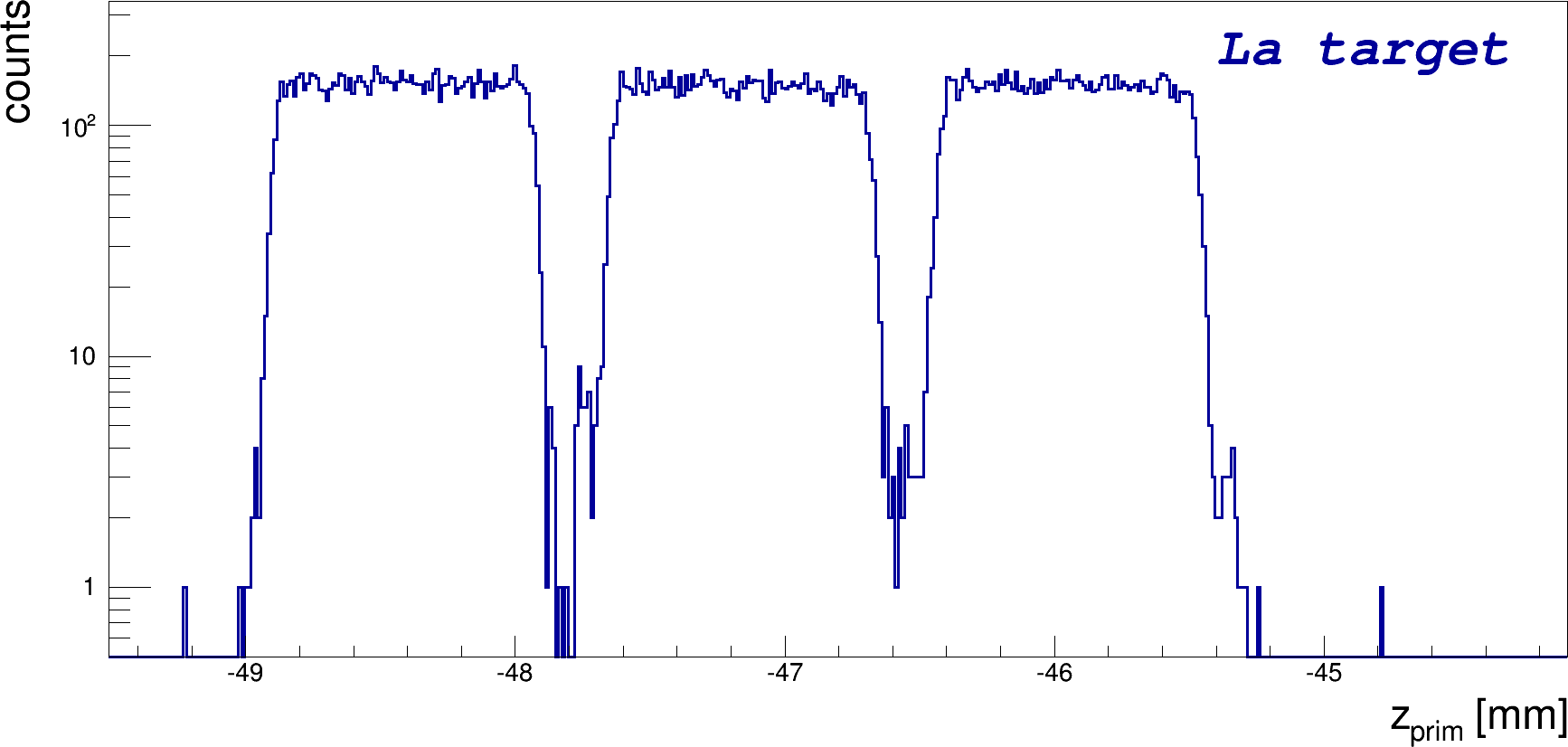} 
\caption{
Distribution of longitudinal coordinate of the primary vertex $z_{prim}$ for interactions in the La target, which was composed of three 1~mm plates.
} 
\label{fig:vd-prim-longitu} 
\end{figure}

Successful performance of the SAVD in 2016 led to the decision to also use it during the Xe+La data taking in 2017.
About $5\cdot10^6$ events of central Xe+La collisions at 150\AGeVc were collected in October and November 2017.
During these measurements the thresholds of the MIMOSA-26 sensors were tuned to obtain high hit detection efficiency which led to significant improvement in the primary vertex reconstruction precision, namely the spatial resolution of the primary vertices obtained for Xe+La data is on the level of $1 ~\mu m$ and $15 ~\mu m$ in the transverse and longitudinal coordinates, respectively.
The distribution of the longitudinal coordinate ($z_{prim}$) of the primary vertex is shown in Figure \ref{fig:vd-prim-longitu} (\textit{left}) (see ~\cite{ana1} for details).
The Xe+La data are currently under analysis and are expected to lead to physics results in the coming months.

The SAVD will also be used during three weeks of Pb+Pb data taking in 2018.
About $1\cdot10^7$ central collisions should be recorded and 2500 \Dzero and \Dzerobar decays can be expected to be reconstructed in this data set.

\section{Proposed measurements after Long Shutdown 2}
\label{sec4}


During the Long Shutdown 2 (LS2) at CERN (2019-2020), a significant modification of the NA61/SHINE spectrometer is planned.
The upgrade is primarily motivated by the charm program which requires a tenfold increase of the data taking rate to about 1~kHz and an increase of the phase-space coverage of the Vertex Detector by a factor of about 2. 
This, in particular, requires construction of the Vertex Detector (VD), replacement of the TPC read-out electronics, implementation of new trigger and data acquisition systems and upgrade of the Projectile Spectator Detector.
Finally, new ToF detectors are planned to be constructed for particle identification at mid-rapidity.
This is mainly motivated by possible future measurements related to the onset of fireball.

The detector upgrades are discussed in detail in \cite{ana1}.
The data taking plan related to the open charm measurements covers measurements of 500M inelastic Pb+Pb collisions at 150\AGeVc in 2021--2024.
This data will provide the mean number of \ccbar pairs in central Pb+Pb collisions needed to investigate the mechanism of charm production in this reaction.
Moreover, the data will allow to establish the centrality dependence of \accbar in Pb+Pb collisions at 150\AGeVc and thus address the question of how the formation of QGP impacts $ \text{J}/\psi $ production.  
Table~\ref{tab:centrality} lists the expected number of charm mesons in centrality selected Pb+Pb collisions at 150\AGeVc assuming the mentioned above statistics of minimum bias collisions.
The estimate is performed assuming that mean multiplicity of charm hadrons is proportional to the number of collisions and using yields calculated for central Pb+Pb collisions within the HSD model~\cite{Linnyk:2008hp,TSong}.
Central (0-30\%) Pb+Pb collisions at 40\AGeVc are planned to be recorded in 2024.
This data together with the result for central Pb+Pb collisions at 150\AGeVc will start a long-term effort to establish the collision energy dependence of \accbar and address the question of how the onset of deconfinement impacts charm production.

\begin{table}[]
\centering
\begin{tabular}{c c c c c c c }
&   0--10\%  & 10--20\% & 20--30\%  &  30--60\%   &  60--90\%  &  0--90\%   \\ \hline 
\#(\Dzero+ \Dzerobar)      &    31k    &   20k     &    11k    &  13k        &   1.3k       &  76k      \\
\#(\Dplus+ \Dminus)        &    19k    &   12k     &    7k    &    8k        &   0.8k     &    46k      \\
$\langle W \rangle$        &    327    &   226     &   156     &  70         &   11       &  105       \\
$\langle N_{COLL} \rangle$   &    749    &   499     &           & 102         &   11       &  202       \\
\end{tabular}
\caption{		
Expected number of charm mesons in centrality selected Pb+Pb collisions at 150\AGeVc assuming 500M minimum bias events recorded in 2022 and 2023, see text for detail.
The mean number of wounded nucleons $\langle W \rangle$ calculated within the Wounded Nucleon Model is also given as well as number of binary collisions.
}
\label{tab:centrality}
\end{table}
The expected high statistics of reconstructed \Dzero and \Dzerobar decays is due to the high event rate and the relatively large efficiencies of open charm detection in the VD. 
The efficiency will be about 13\% (3 times better than for the SAVD) for the \Dzero  $\rightarrow \pip + \km$ decay channel and about 9\% \footnote{The quoted efficiencies include the geometrical acceptance for \Dzero $\rightarrow \pip + \km$ (\Dplus $\rightarrow \pip + \pip + \km$) decays and the efficiency of the analysis quality cuts used to reduced the combinatorial background.} for \Dplus decaying into $\pip + \pip + \km$. 

\begin{figure}[]
\centering
\includegraphics[width=0.65\textwidth]{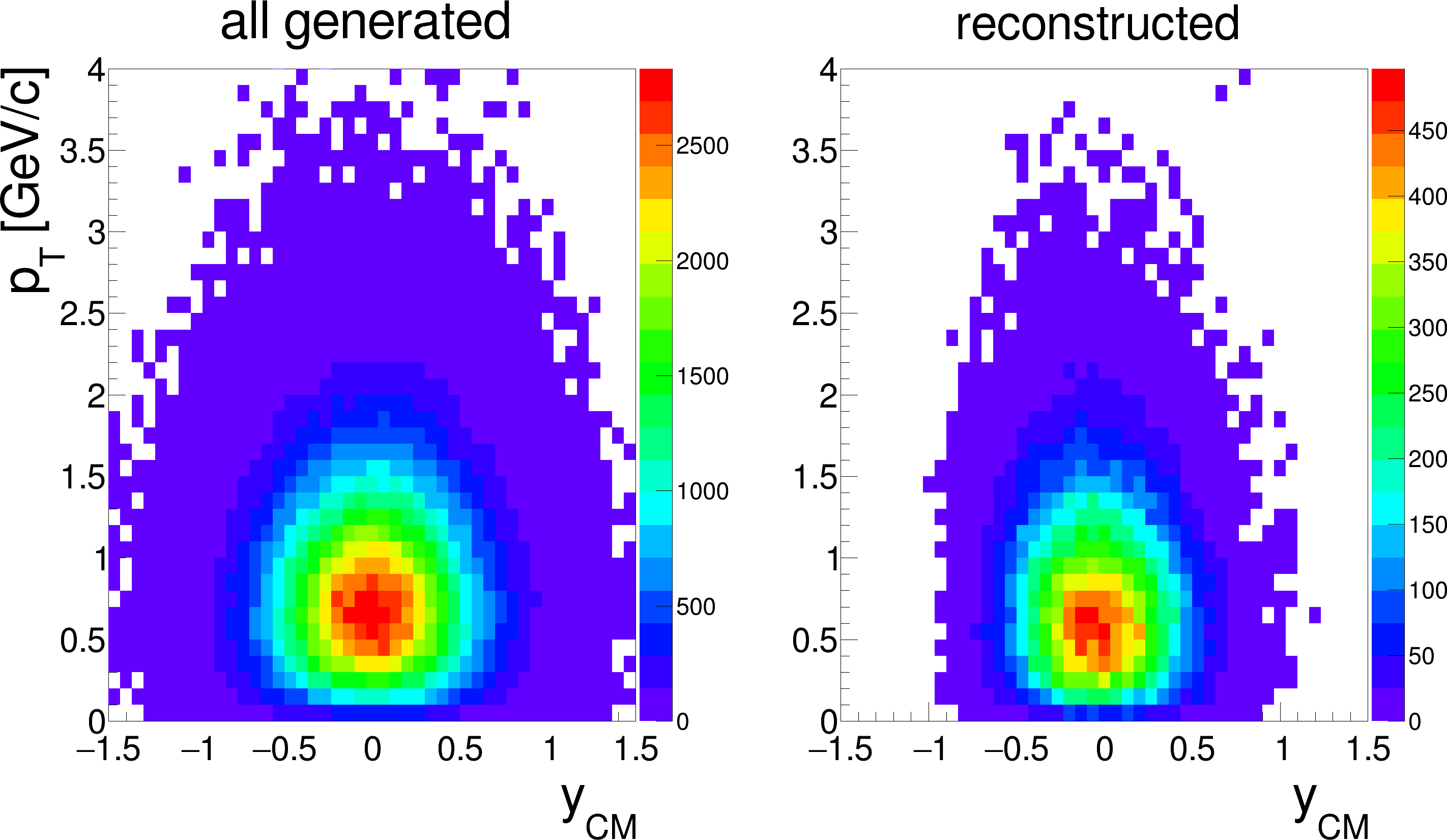}
\caption{
Transverse momentum and rapidity distribution of \Dzero+ \Dzerobar mesons produced in about 500M inelastic Pb+Pb collisions at 150\AGeVc for all produced \Dzero+ \Dzerobar mesons (\textit{left}) and  \Dzero+ \Dzerobar mesons fulfilling the following criteria: decay $\Dzero \to \pi^+ + \km$, decay products registered by the VD, passing background suppression cuts \textit{(right)}.}
\label{fig:acceptSimLAVD}
\end{figure}

\begin{figure}[]
\centering
\includegraphics[width=0.85\textwidth]{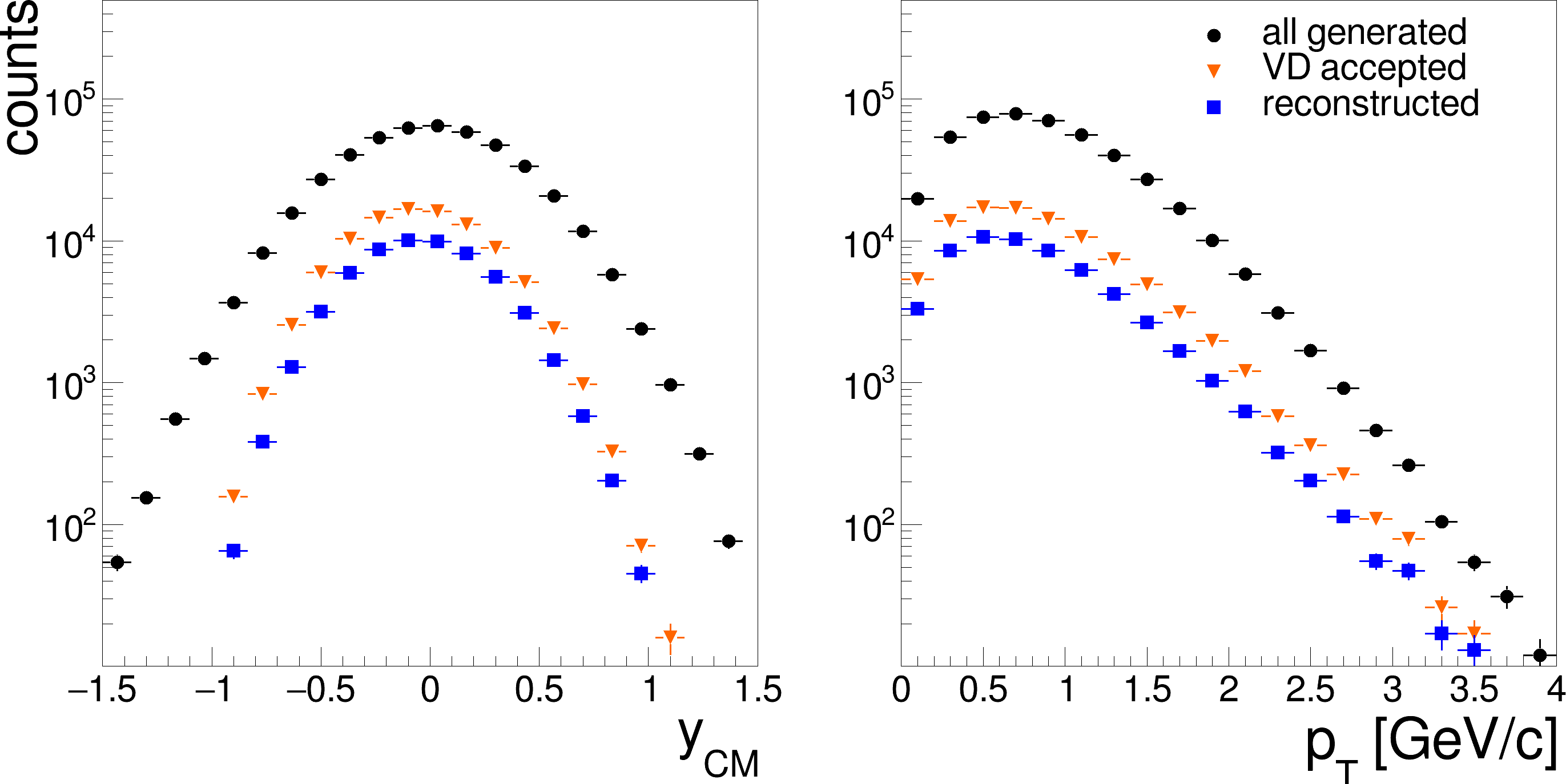}
\caption{
Rapidity (\textit{left}) and  transverse momentum (\textit{right}) distributions of \Dzero+ \Dzerobar mesons produced in about 500M inelastic Pb+Pb collisions at 150\AGeVc.
Dots indicate all generated mesons, triangles mesons within the VD acceptance and squares mesons within the VD acceptance and passing background suppression cuts.
}
\label{fig:y:full}
\end{figure}

Figure~\ref{fig:fig:y:full} shows distributions of \Dzero + \Dzerobar mesons in rapidity and transverse momentum for all generated particles (black symbols)  and for particles that passed the acceptance and background reduction cuts (blue symbols).
The presented plots refer to 500M inelastic Pb+Pb collisions at 150\AGeVc. 
Based on the presented simulations one estimates, that fully corrected results will correspond to more than 90\% of the \Dzero and \Dzerobar yield (see Figs.~\ref{fig:acceptSimLAVD} and~\ref{fig:y:full}).
Total uncertainty of $ \langle \Dzero \rangle $ and $ \langle \Dzerobar \rangle $ is expected to be about 10\% and is dominated by systematic uncertainty.

In summary it is emphasised that only NA61/SHINE is able to measure open charm production in heavy ion collisions in full phase space and in the beginning of the next decade. 
The corresponding potential measurements at higher (LHC, RHIC) and lower (FAIR, J-PARC) energies are necessary to complement the NA61/SHINE results and establish collision energy dependence of charm production.

\section*{Acknowledgements}
This work was supported by the Polish National Centre for Science grants 2014/15/B/ST2/02537 and 2015/18/M/ST2/00125. 

\bibliographystyle{elsarticle-num}

\end{document}